\documentclass[preprint,showpacs,preprintnumbers,amsmath,amssymb]{revtex4}

\usepackage{graphicx}
\usepackage{dcolumn}
\usepackage{bm}

\begin{document}

\title{Fast Quantum Computing with Buckyballs}

\author{Maria Silvia Garelli and  Feodor V Kusmartsev }
\date{18.07.2005}

\affiliation{Department of Physics, Loughborough University, LE11 3TU, UK}

\begin{abstract}
We have found that encapsulated atoms in fullerene molecules, which carry a spin, can be used for fast quantum computing. We describe the scheme for performing quantum computations, going through the preparation of the qubit state and the realization of a two-qubit quantum gate. When we apply a static magnetic field to each encased spin, we find out the ideal design for the preparation of the quantum state. Therefore, adding to our system a time dependent magnetic field, we can perform a phase-gate. The operational time related to a $\pi-$phase gate is of the order of $ns$. This finding shows that, during the decoherence time, which is proportional to $\mu s$, we can perform many thousands of gate operations. In addition, the two-qubit state which arises after a $\pi-$gate is characterized by a high degree of entanglement. This opens a new avenue for the implementation of fast quantum computation. 

\end{abstract}


\maketitle
\section{Introduction}

In the study of systems for the realization of quantum gates, a great interest is addressed to encoding qubits in spins. The most remarkable property of both nuclear and electronic spin is their long decoherence time, which allows them to be the most promising objects for quantum manipulations. The first studies about quantum computing via spin-spin interaction were based on the Nuclear Magnetic Resonance (NMR) technique \cite{Gershenfeld,Schmidt,Leibfried,Nielsen},
but these systems show a very limited scalability in the number of qubits. 
Solid state devices were found to be suitable for building up scalable spin-based quantum computers \cite{Kane,Loss,Div}.
In this work we focus on the realization of a spin based quantum gate, considering a system composed of endohedral fullerene molecules. The qubit is encoded in the electron spin of the encased atom in each fullerene molecule. Many studies about the physics of endohedrally doped fullerenes have been performed \cite{Harneit,Harneit1,Feng}. In our study we borrow many ideas from these previous papers, but we use a different approach for performing the gate operation and entangled states. Our main target is to get a very short operational time for such tasks.  
The chemical and physical properties of an endohedral fullerene molecule, see Refs. \cite{Greer,Harneit,Suter,Twam}, are very remarkable. Any charge inside these molecules is completely screened, and the fullerene can be considered as a Faraday cage which traps the encased atom. Moreover, considering the $N@C_{60}$, the nitrogen atom sits in the center of the fullerene cage and it preserves all the characteristics of the free nitrogen together with a lower reactivity. Indeed, this buckyball is stable also at room temperature. In addition, the mutual interaction between two spins in adjacent buckyballs, is dominated by the spin dipole-dipole interaction, while the exchange interaction vanishes. The most relevant feature which is required for a reliable quantum computation, is the long decoherence time of spins trapped in fullerenes.

These endohedral systems are
typically characterized by two relaxation times. The first is
$T_1$, which is due to the interactions between a spin and the
surrounding environment. The second relaxation time is $T_2$ and it is due to
the dipolar interaction between the qubit encoding spin and the
surrounding endohedral spins randomly distributed in the sample. While
$T_1$ is dependent on temperature, $T_2$ is practically
independent of it. The experimental measure of the two relaxation
times shows that $T_1$ increases with decreasing temperature from
about $100\mu s$ at $T=300 K$ to several seconds below $T=5K$, and
that the value of the other relaxation time, $T_2$, remains constant, that is  $T_2\simeq 20\mu s$ \cite{Knorr1,Knorr2}. It is thought that the value of $T_2$ can be
increased, if it is possible to design a careful experimental
architecture, which could screen the interaction of the spins with the
surrounding magnetic moments. 
Actually, the \emph{peapod} is the most promising setup for trapping buckyballs \cite{Briggs1}, and for getting better decoherence times \cite{Briggs,site1,site2}. In such a peapod, different ordered phases of fullerenes, which could be relevant to quantum computation, have been observed \cite{Briggs2}.

We studied the realization of a fast quantum computation. We focused on the implementation of a two-qubit
quantum \emph{$\pi$-gate}, which is a generalization of the
\emph{phase gate}. The theoretical study related to the realization of the $\pi-$gate is treated in our previous work, \cite{Garelli}. In contrast with our previous work here we found a new setup which allows us to perform many quantum operations characterized by a very short operational time. We show that during such short time we were able to create a two-qubit operation like a $\pi-$gate and a highly entangled state. We have also designed the ideal preparatory and final setup needed before and after the realization a two-qubit gate.
 In this setup, before performing the two-qubit operation, we initially apply a static magnetic field in the z-direction in order to lift the energy degeneracy on each of the two spin$-\frac{1}{2}$ particles. Moreover, this design is suitable as a starting configuration for the following two-qubit operation. This configuration for the system can be also adopted at the end of the gate operation, in order to preserve the result obtained through its computation. To perform the gate, we apply an additional microwave magnetic field, always in the z-direction. The spin dipole-dipole interaction of the two-qubit system, controlled by the added static and microwave fields, enables the system to realize the desirable $\pi-$gate. As we introduced before, the main result of our study
is the gate time, i.e. the time required by the system in order
to complete a $\pi$-gate operation. The value obtained for the gate time through numerical computation is
$\tau\simeq1.6 ns$, which is about three orders smaller than
the shortest relaxation time, $T_2$ \cite{Knorr1,Knorr2}. Comparing $\tau$ and $T_2$, we found that it
is theoretically possible to realize many thousands of basic gate
operations before the system decoheres. We also found that the two-qubit state arising during the gate operation is characterized by a high degree of entanglement. Evaluating the \emph{concurrence} of the two-qubit state, see Ref. \cite{Wootters}, we see that, at the gate time, $\tau$, the corresponding value of concurrence is $C\simeq 0.9$. This value is close to the maximum value for the concurrence, i.e. $C=1$, therefore our two-qubit state reaches a good degree of entanglement at the end of the $\pi-$gate operation.

Therefore, we can perform a $\pi-$gate in a remarkable short time, and, during this time interval, the two- qubit state acquires a high level of entanglement, which allows it to be a good candidate for carrying quantum information.  
\section{Realization of a $\pi-$gate}

Our system consists of two spins, which interact with a static
magnetic field. By applying a static magnetic field oriented in the
$z$ direction we get splitting of the
spin z component into the spin-up and spin-down components, which is due to the Zeeman effect. The
energy difference between these two levels give the resonance
frequency of the particle. If we apply a static magnetic
field to the whole sample, all the particles will have the same
resonance frequency. If we put these buckyballs in a frequency resonator, each buckyball must have its own resonance frequency, in order to be individually addressed and manipulated. This arrangement leads to the most relevant
experimental disadvantage for systems composed by arrays of
buckyballs, which is the difficulty in the individual addressing
of each qubit. Although the single addressing of each qubit is not strictly relevant for the realization of our two-qubit gate, it is useful however for performing single qubit operations. Indeed, for the realization of this type of operation, we need to be able to distinguish qubits, in order to act on them independently.

In order to overcome the problem of addressing a single qubit, external field gradients, which
can shift the electronic resonance frequency of the qubit-encoding
spins, can be used \cite{Suter}. With the use of atom chips, thin wires
can carry a current density of more than $10^7 A/cm^2$ \cite{Groth}, therefore we can realize the desirable magnetic field gradients \cite{Harneit1,Garelli}. 
The values of the arising magnetic field amplitudes on our two spins are $B_{g_1}=3.73\times 10^{-5}T$ and $B_{g_1}=-3.73\times 10^{-5}T$, for the left and right spin respectively.

Choosing a static magnetic field along the z direction, and neglecting the exchange interaction between the two spins \cite{Greer,Waiblinger,Harneit}, the Hamiltonian of our system is the following 
\begin{equation}\label{hamtind}
\begin{array}{ll}
H&=g(r)[\vec{\hat\sigma}_1\cdot
\vec{\hat\sigma}_2-3(\vec{\hat\sigma}_1\cdot\vec n)(\vec{\hat\sigma}_2\cdot\vec
n)]\\
&-\mu_B[((B_{z}+B_{g_1})\hat\sigma_{z_1})\otimes
I_2\\
&+I_1\otimes((B_{z}+B_{g_2})\hat\sigma_{z_2})],
\end{array}
\end{equation}
where $g(r)=\gamma_1 \gamma_2 \frac{\mu_0 \mu_{B}^{2}}{4 \pi r^3}$, $\mu_0$ is the diamagnetic constant, $\mu_B$ is the Bohr magneton and $\vec{\hat\sigma}_{1,2}$ are the spin matrices. Choosing $\gamma_{1}=\gamma_2=2$ for the gyromagnetic ratio, we obtain $g(r)=\frac{\mu_0 \mu_{B}^{2}}{\pi r^3}$.
Whenever we are performing a quantum gate, we have to be able to stop the gate operation. Since our gate is due to the time evolution of the system, we have to find a way to slow down the interaction process after a $\pi-$gate is performed. Static magnetic fields cannot be switched off immediately. Because of the circuit inductance, the static field vanishes slowly in comparison to the gate time we are looking for. On the other hand, time dependent oscillating fields, like microwave fields, can be switched off promptly. Therefore, we have to look for the best microwave field, which will dominate the time evolution of the quantum state of the system. 
\subsection{Preparatory Configuration}

\begin{figure}[htb]
\centering
\includegraphics[scale=0.8]{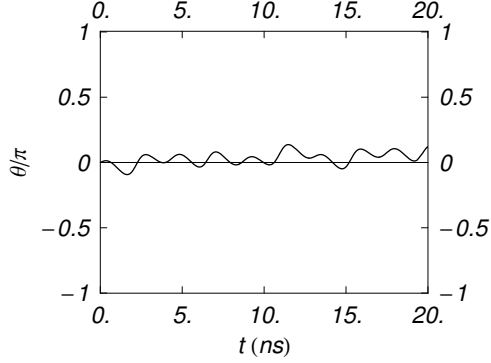}
\caption{Evolution of the phase $\theta$ during time, $t$. The two-spin system is subjected to a static field $B_z$ and to the two amplitudes $B_{g_1}$ and $B_{g_2}$, due to the field gradient. The time range is up to $20 ns$.}\label{static}
\end{figure}

\begin{figure}[htb]
\centering
\includegraphics[scale=0.8]{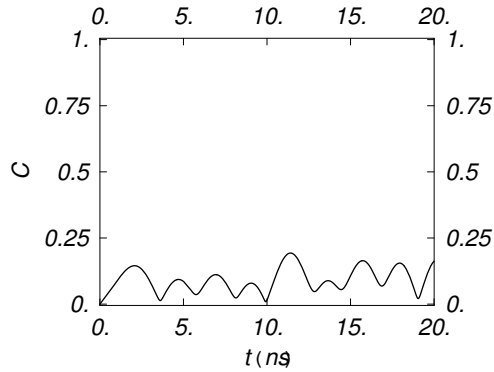}
\caption{Dependence of concurrence, $C$, on time, $t$. The time range is up to $20 ns$.}\label{ctind}
\end{figure}

At first, we consider the case where, in addition to terms $B_{g_1}$ and $B_{g_2}$, we only apply a static magnetic field, $B_z$, in the z direction to our system. To check the time evolution of the phase $\theta$ we need to solve the Schr{\"o}dinger equation, in which the Hamiltonian is given by Eq. (\ref{hamtind}), and the time evolved wave function of the system is
\begin{equation}\label{wave}
\begin{array}{lll}
\mid \psi(t)\rangle&=c1(t)\mid 00\rangle+c2(t)\mid 01\rangle\\
&+c3(t)\mid 10\rangle+c4(t)\mid 11\rangle.
\end{array}
\end{equation}
Therefore, we get the following four differential equation system
\begin{eqnarray}\label{systemTindip1}
\dot c1(t)&=&-\imath[(g(r)+m_1)c1(t)-3g(r)c4(t)];\\
\dot c2(t)&=&-\imath[(-g(r)+m_2)c2(t)-g(r)c3(t)];\\
\dot c3(t)&=&-\imath[
-g(r)c2(t)+(-g(r)-m_2)c3(t)];\\\label{systemTindip4} \dot
c4(t)&=&-\imath[-3g(r) c1(t)+(g(r)-m_1)c4(t)],
\end{eqnarray}
whose solution gives the phases acquired by each computational basis state during the time evolution. According to the formula which allows us to evaluate a phase-gate, see Ref. \cite{Garelli}, we can evaluate phase $\theta$ as follows
\begin{equation}\label{theta1}
\begin{array}{ll}
\theta &=Arg(c1(t))-Arg(c2(t))\\
&-Arg(c3(t))+Arg(c4(t)),
\end{array}
\end{equation}
where $Arg(c(i))$, $i=1,..4,$, are the phases of the complex coefficients related to each basis state.
 Choosing $B_{z}=5\times 10^{-4}T$, we obtain the result of the numerical computation for phase $\theta$, see Fig. \ref{static}. 
In this picture the time interval is up to $t=20 ns$, and we can see that in this time range the phase $\theta$ shows small oscillations about zero and it is very far from $\pi$. These phase oscillations are negligible in comparison to a $\pi-$gate, and we could consider the behaviour phase $\theta$ as approximately constant during this time. We also checked the concurrence of the two qubit state. Using the following formula
 \begin{equation}\label{concnorm}
C(\psi)=\frac{2\mid c2^*c3^*-c1^*c4^*\mid}{\mid c1\mid^2+\mid
c3\mid^2+\mid c3\mid^2+\mid c4\mid^2}
\end{equation}
whose derivation can be found in Ref. \cite{Garelli} , we performed a numerical evaluation of the concurrence. The concurrence is a positive monotonically increasing time dependent function, which shows the degree of entanglement of a qubit state. When concurrence is minimum, $C=0$, the related quantum state is separable. When it reaches its maximum value, $C=1$, the related state is maximally entangled. If each qubit in a maximally entangled state undergoes a spin-flip transformation, the total state does not change at all, therefore it is the best candidate for carrying quantum information. In Fig. \ref{ctind}, the time evolution of concurrence in a time interval of $20 ns$ is presented.

We can see that in this time range, not only does the phase $\theta$ shows negligible oscillations, but also the concurrence is characterized by small deviations about zero. Therefore, the concurrence is well-approximated by a constant function nearly equal to zero. This finding enables us to conclude that the two-qubit state obtains a negligible degree of entanglement during the time evolution, when the system is subjected to only the chosen static field $B_z$.
This setup is very good as a starting and final configuration, corresponding to the state before and after the realization of the $\pi-$gate. Indeed, we need the two-qubit state not to be entangled before the two-qubit gate operation is started. Moreover, this setup could also be applied at the end of the gate operation. Indeed, when the system is subjected to the static field $B_z$, and to the two amplitudes $B_{g_1}$ and $B_{g_2}$, the evolution of phase $\theta$ remains approximately constant and the two-qubit state preserves its degree of entanglement. This is a way to preserve the result obtained through the computation of the two-qubit operation. It is important to note that if the two-buckyball system is allowed to evolve without any applied magnetic field, the interaction is represented by only the mutual dipole-dipole spin interaction. In this arrangement, the time evolution provides a phase $\theta$ constantly equal to zero, therefore the quantum state which describe the system never becomes entangled.  

\subsection{Setup for Performing a $\pi-$gate}
\begin{figure}[htb]
\centering
\includegraphics[scale=0.8]{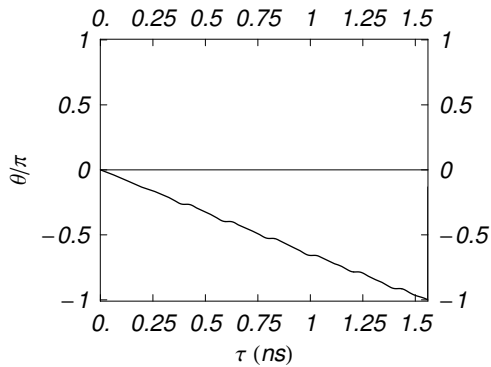}
\caption{$\pi-$gate: time evolution of phase $\theta$. The gate time, corresponding to $\theta=-\pi$, is $\tau\simeq1.56 ns$.}\label{microz}
\end{figure}

We will now show the setup for the realization of the $\pi-$gate, and the way in which we can create an entangled state. We keep the configuration of the previous setup, but we also apply a microwave field or a time oscillating magnetic field oriented in the z direction, of the form $B(t)=B_t \cos (\omega t)$. Therefore, the Hamiltonian of the system is 
\begin{equation}\label{hamtdip}
\begin{array}{ll}
H&=g(r)[\vec{\hat\sigma}_1\cdot
\vec{\hat\sigma}_2-3(\vec{\hat\sigma}_1\cdot\vec n)(\vec{\hat\sigma}_2\cdot\vec
n)]\\
&-\mu_B[((B_{z}(t)+B_{g_1})\hat\sigma_{z_1})\otimes
I_2\\
&+I_1\otimes((B_{z}(t)+B_{g_2})\hat\sigma_{z_2})],
\end{array}
\end{equation}
where $B_{z}(t)=B_z+B(t)$ is the total magnetic field in the z direction. Solving the Schr{\"o}dinger equation with the Hamiltonian given by Eq. (\ref{hamtdip}), the phases related to each basis state are derived as in the time independent case. Manipulating these phases as shown in Ref. \cite{Garelli}, through the requirement 
\begin{equation}
\theta=\pm\pi,
\end{equation}
the gate time (i.e. the time taken by phase $\theta$ to reach a value equal to $-\pi$ or $+\pi$) is evaluated.
After many numerical trials, the optimal choice for the amplitude and the frequency of the time dependent field is $B_t=2\times 10^{-1}T$ and $\omega=15.5 GHz$, respectively. The evolution of $\theta$ is shown in Fig. \ref{microz}. The gate time, corresponding to $\theta=-\pi$, is $\tau=1.56 ns$. In Fig. \ref{microz}, the phase evolution is not smooth, but shows small jumps.

 These small jumps are due to the oscillation period of frequency $\omega$ of the microwave field. 
 
\begin{figure}[htb]
\centering
\includegraphics[scale=0.8]{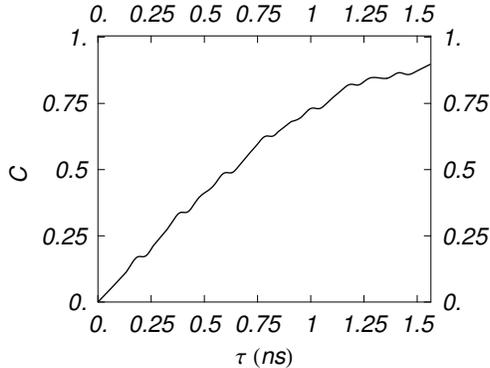}
\caption{Time evolution of the concurrence. At the gate time, $\tau\simeq 1.56 ns$, the concurrence value is $C\simeq0.90$.}\label{cmicroz}
\end{figure}

 The most interesting result is finding the time evolution of the concurrence, see Fig. \ref{cmicroz}.  
 The concurrence is $C(t)=0.90$ at the gate time, $\tau$. Since the concurrence is close to its maximum value, we can say that, at the end of the gate operation, the two-qubit state is strongly entangled. Therefore, the final state of our $\pi-$gate is reliable for carrying quantum information. The time evolution of the concurrence shows small jumps, always due to the oscillation period of $\omega$, as in the case of phase evolution.

\section{Conclusions}
Our study has been focused on the realization of a high speed $\pi-$gate, which is a particular choice of a two-qubit phase gate. We have also looked for the best configuration of the system, before and after the implementation of the $\pi-$gate. 
Our quantum computation is performed through the evolution of a system composed by two spin-$\frac{1}{2}$ particles. 

First, we have investigated the optimal setup for the preparatory configuration. To encode the qubit in each spin, we applied  static magnetic fields oriented in the z direction. This procedure causes each spin to undergo the Zeeman splitting of the spin z-component. The two-level systems, which arise as a result, represent individual qubits. By applying small amplitude magnetic fields, as discussed in the previous section, we showed that the concurrence value of the two-qubit preparatory state is approximately $C\simeq 0$, see Fig. \ref{ctind}. This result shows that our preparatory state does not become entangled when it is subjected only to static fields, at least for the characteristic time of interest. However, it could get entangled after a very long time. Therefore, we can assume that the starting state in the computation of the $\pi-$gate is approximately not entangled.  This is the ideal initial condition for the implementation of the two-qubit gate.

Next, we investigated the realization of a quantum phase-gate. The phase gate is performed simply by switching on a microwave field oriented in the z direction, which has some optimal amplitude. Since the time dependent field can be turned on and off promptly, the  phase gate operation can be controlled and stopped at any time. Therefore, any phase gate can be realized, i.e. $\pm\frac{\pi}{4}$, $\pm\frac{\pi}{2}$, $\pm\pi$. The only quantity of interest in the computation of a phase gate is the phase $\theta$, see Eq. \ref{theta1}. For the realization of the $\pi-$gate, we require $\theta=-\pi$. The action of only static fields on the system results in a practically constant phase $\theta$, see Fig. \ref{static}. Therefore the implementation of the $\pi-$gate is limited to the time range between the switching on and off of the time dependent magnetic field. The operational time needed to perform the $\pi-$gate is $\tau\simeq 1.6 ns$,which is remarkably short. 

At the end of the gate operation we should be able to preserve its result, that is $\theta=-\pi$. When we turn off the time dependent field and allow the system to evolve only under the action of static fields, the phase $\theta$ shows again a constant behaviour, as it was before the phase gate was applied. This is a convenient and reliable arrangement to preserve the final result obtained in the gate operation and serves as a good preparatory state for the next quantum computation. Moreover, this set up enables the two-qubit state to retain its characteristics that it obtained at the end of the gate operation. Since the $\pi-$gate operation enables the two-qubit state to become strongly entangled, see Fig. \ref{microz}, by switching off the time dependent field and subjecting it to the same static fields only, the state will preserve in time its degree of entanglement.  

Finally, our scheme for a quantum computation will allow the system, consisting of two spins, to perform a fast two qubit gate and to create a very robust entangled state. If we have a system with more than two spins, we can create an entangled state between all spins by applying this two-qubit gate in sequence to any pair of spins. Furthermore, such state could be used in other quantum computations, i.e. quantum error correction codes.
\section{Acknowledgments}
The authors are grateful and thankful to Andrew Briggs and Jason Twamley for their very helpful discussions. Many thanks to Debbie Dalton and Neil Lindsey for their help with corrections and computer related troubles. MSG is pleased to thank Giuseppe Giordano for his never-ending and much valuable moral support and friendship.

\end{document}